\documentclass[]{spie}  

\usepackage{amsmath,amsfonts,amssymb}
\usepackage{graphicx}
\usepackage[colorlinks=true, allcolors=blue]{hyperref}

\title{Project management of the Canadian Hydrogen Observatory and Radio-transient Detector}

\author[a,b]{Dallas Wulf}
\author[a,b]{Matt Dobbs}
\author[c]{Brian Hoff}
\author[c]{Michael Rupen}
\author[d]{Kendrick Smith}
\author[e,f]{Keith Vanderlinde}
\affil[a]{Department of Physics, McGill University, Montr\'eal, QC, Canada}
\affil[b]{Trottier Space Institute, McGill University, Montr\'eal, QC, Canada}
\affil[c]{Dominion Radio Astrophysical Observatory, Herzeberg Astronomy \& Astrophysics Research Centre, National Research Council Canada, Kaleden, BC Canada}
\affil[d]{Perimeter Institute for Theoretical Physics, Waterloo, ON, Canada}
\affil[e]{David A.\ Dunlap Department of Astronomy \& Astrophysics, University of Toronto, Toronto, ON, Canada}
\affil[f]{Dunlap Institute for Astronomy and Astrophysics, University of Toronto, Toronto, ON, Canada}

\authorinfo{Send correspondence to D.W. at dallas.wulf@mcgill.ca}

\begin{document} 
\maketitle

\begin{abstract}
The Canadian Hydrogen Observatory and Radio-transient Detector (CHORD) is a new drift-scan radio interferometer operating from 300–1500 MHz, optimized for 21cm intensity mapping, 21cm galaxy detection, pulsars, and transient detection and localization. The full instrument will feature a core array of 512 six-meter dishes, plus two 64-dish outriggers at several thousand kilometer baselines. The project is nearing completion of the first 64-dish sub-array, which will enable commissioning activities to begin in parallel with the remaining construction. In this presentation, we discuss the project management framework that has enabled us to navigate the logistical challenges of a fixed-budget, fixed-timeline research infrastructure project through a period of significant global economic and technological volatility.
\end{abstract}

\keywords{Project Management, Facilities, Observatories, Radio Telescopes, Interferometers}

\section{INTRODUCTION}
\label{sec:intro}  

The Canadian Hydrogen Observatory and Radio-transient Detector (CHORD) is a new drift-scan radio interferometer under construction at the Dominion Radio Astrophysical Observatory near Penticton, British Columbia, Canada.  When completed, the core instrument will consist of 512 6-meter dishes arranged in a close-packed grid to optimize redundancy and mapping speed. There are also two outrigger stations planned, each consisting of 64 6-meter dishes, at the Green Bank Observatory in West Virginia, USA and the Hat Creek Observatory in California, USA. CHORD’s combination of high sensitivity and low systematics at the front end, coupled with state-of-the-art hardware and software in the correlator will enable world-leading capabilities in the areas of 21cm intensity mapping, 21cm galaxy detection, pulsars, and transient detection and localization. For a more in-depth description of CHORD, please refer to Ref.~\citenum{Dobbs2026}.

CHORD is primarily funded by the Canada Foundation for Innovation (CFI), which is the primary agency for funding research infrastructure projects in Canada. The proposal to fund CHORD was submitted in 2019, though the project did not start in earnest until late 2022, due to administrative delays related to the COVID-19 pandemic. Funding is administered through McGill University, the University of Toronto, the University of Calgary, the Perimeter Institute, and the National Research Council of Canada.  CHORD has also received significant funding for its GPU correlator from Italy’s Istituto Nazionale di Astrofisica (INAF). From these founding partners, the collaboration has grown to 95 active members from 14 institutions, including faculty, staff, and trainees.  The collaboration is organized into several working groups, each addressing a particular subsystem of the instrument or scientific objective (see Fig.~\ref{fig:org}).

\section{PROJECT MANAGEMENT IN CHORD}
\label{sec:pm}

By design, CFI-funded projects are selected to have sufficiently high technical readiness, such that they can be completed and have significant impact in a relatively short time frame (typically 3-5 years, depending on the size of the award) \cite{noauthor_innovation_2024}.  A high level of technical readiness also means that these projects are quite constrained in terms of having fixed budgets, timelines, and outcomes. Nevertheless, the recent pace of economic, political, and technological change still poses a significant challenge to managing these projects.  As an example, consider the computing industry: the availability of many computer components cannot be guaranteed three years into the future, and even when items remain on the market, their costs can easily change by a factor of a few (as demonstrated by RAM in recent years).  As a result, the hardware and its capabilities that were assumed in the original project plan are often different from what is ultimately implemented. Only the most high-level project requirements can be adhered to, and even then, some tradeoffs are likely if not inevitable to meet project constraints. 

Software and tech industries have long recognized challenges such as these, and responded by developing so-called “agile” approaches to management \cite{white_agile_2008}. Agile management approaches, among other things, place less emphasis on pre-defined requirements and planning, and more emphasis on being able to adapt quickly in the presence of large uncertainties.  It is therefore unsurprising that our approach to project management within CHORD features several of the hallmarks of these agile techniques, such as:
\begin{itemize}
  \item Close interaction between instrument builders/developers and end users;
  \item Small, highly skilled teams;
  \item Adaptive leadership, empowering subject matter experts to make decisions; 
  \item Iterative approaches to defining requirements; and
  \item Delayed decision making and just-in-time methodologies.
\end{itemize}

\begin{figure} [ht]
   \begin{center}
   \begin{tabular}{c} 
   \includegraphics[width=15cm]{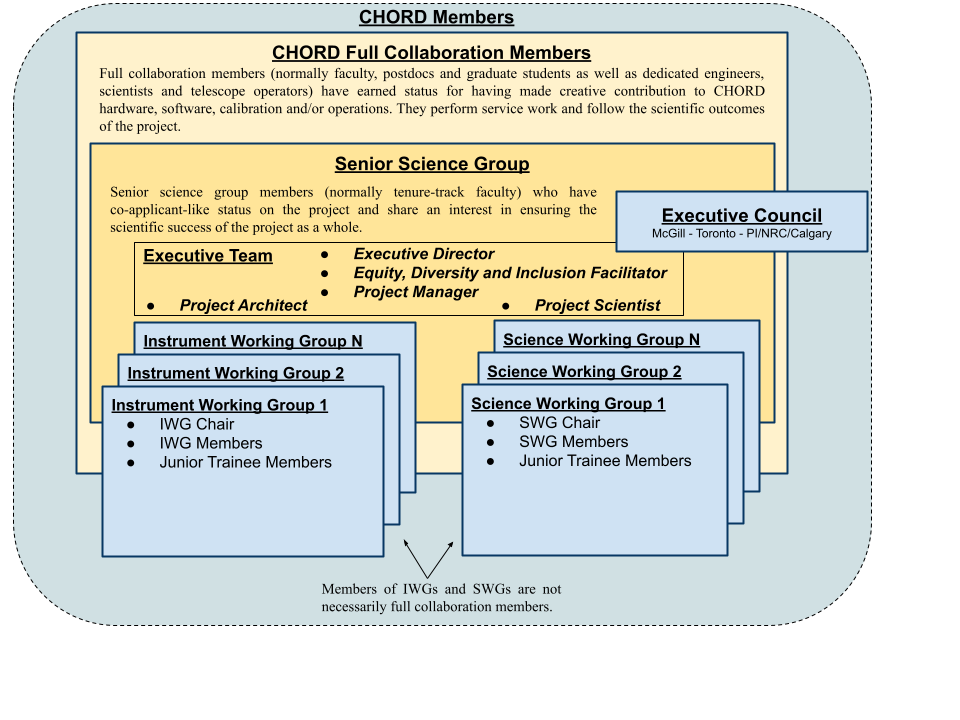}
   \end{tabular}
   \end{center}
   \caption[example] 
   { \label{fig:org} 
The CHORD collaboration organization structure reproduced from the CHORD collaboration agreement. Each instrument subsystem and science case is organized into a working group. The Senior Science Group and Executive Team is responsible for ensuring that the high level requirements of CHORD are met, while leaving the low level requirements and implementation details to the working groups, which hold the subject matter expertise. This structure contributes to CHORD's agility and resilience to uncertainty.}
   \end{figure}

\begin{figure} [ht]
   \begin{center}
   \begin{tabular}{c} 
   \includegraphics[width=15cm]{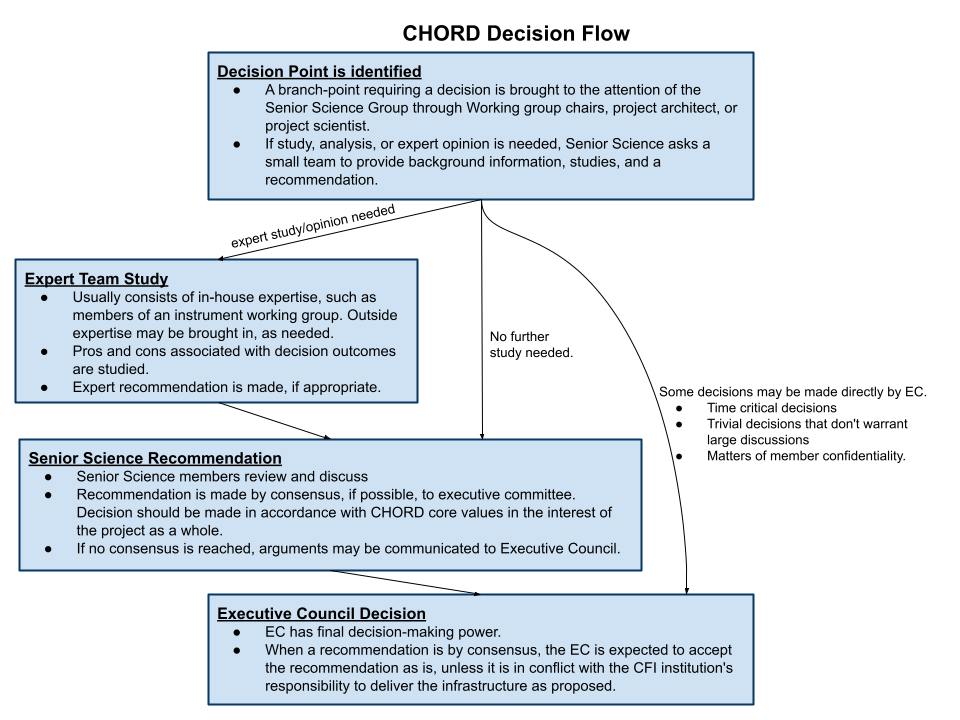}
   \end{tabular}
   \end{center}
   \caption[example] 
   { \label{fig:decision} 
The CHORD decision making process reproduced from the CHORD collaboration agreement, emphasizing the role of small, expert teams. In most cases, decisions reflect the recommendation of these expert teams. Senior Science and Executive Council oversight only serves to ensure that decisions are consistent with CHORD's high level objectives and administrative requirements.}
   \end{figure}

\noindent As an example of these techniques in practice, we show the CHORD collaboration organization structure and decision flow in Figs.~\ref{fig:org} and \ref{fig:decision}, respectively.  

Anecdotally, these techniques have served CHORD well, or at the very least CHORD has progressed well in spite of them. Admittedly, the dynamism that results from this approach can sometimes border on chaos, or amplify risks that would be mitigated under more traditional approaches to project management. Nevertheless, four years since its inception, CHORD has experienced only minor schedule slips (with the exception of early pandemic-related delays) and is on track to deliver the performance forecasted in the original proposal \cite{vanderlinde_lrp_2019}.  After demonstrating first fringes in the Fall of 2025, CHORD is currently working to complete the "pathfinder" phase of the project, which comprises the first 64-dish sub-array.  This phase of the project will enable the start of commissioning activities and even some early science observations in parallel with the completion of the core array and outriggers.

\section{CHALLENGES}
\label{sec:challenge}

Despite CHORD’s successes, there have also been challenges.  Perhaps unsurprisingly, the most significant challenges---and the bulk of CHORD's project management effort---relate to issues with procurement and managing the budget. After all, about 80\% of the budget for infrastructure projects such as CHORD is spent in the form of purchase orders (PO) for goods and services (the balance is spent on staff salaries).  Moreover, as CHORD operates as its own general contractor, these expenses are spread over hundreds of transactions, as opposed to a few big ticket purchases. Ensuring that these procurements meet all of our technical requirements while staying on schedule and on budget requires constant effort. In the following sections, we highlight some of these challenges and how they are managed within CHORD.
 
\subsection{Competitive Procurement}
\label{sec:procure}

Most public institutions implement some form of competitive procurement strategy---ranging from simple quotes for low-value items to public tenders for large purchases---to select vendors and products in a manner that strives to ensure fairness while maximizing value. For common goods and services, the linear process of defining requirements and then evaluating quotes or bids works reasonably well to achieve the intended objectives. 

However, this process presents challenges for more specialized purchases.  First, there are cases where technical specifications depend on one another (e.g. specification A depends on specification B), which can make it difficult to clearly communicate requirements to vendors.  Second, there is the challenge of establishing quantitative evaluation criteria, which requires anticipating potential tradeoffs between seemingly unrelated variables (e.g. balancing longer warranty against better performance). 

The issue is not a lack of technical knowledge, but the assumption that such complex requirements and evaluation criteria can be written \textit{a priori} and in a way that is vendor agnostic. Much like defining project requirements in an agile framework, defining complex procurement requirements is inherently non-linear. In our experience, it requires an iterative approach and early consultation with potential vendors to align requirements with industry capabilities. Consequently, by the time CHORD issues a formal public tender, teams have already explored the trade-offs and identified a preferred solution, which naturally shapes the final criteria.

Despite these challenges, CHORD has been able to implement this iterative process while remaining compliant with institutional directives. The primary drawback with this approach is administrative bloat that occurs from effectively running two procurement processes in series---our own internal iteration, followed by the institutionally mandated process. Additionally, we have yet to be able to enforce a uniform approach to this iterative process across CHORD's multiple instrument working groups. For example, risk tolerance varies between groups, both due to differences in personalities and fundamental differences in how risk affects different subsystems.  In any case, some working groups are more willing than others to explore new vendors or relax requirements in the face changing costs or availability. This has created some additional project management challenges to maintain budgets and deliver uniform outcomes, and remains an area of active development and improvement.

\subsection{Procurement Risk Management}
\label{sec:risk}

Keeping to the procurement theme, another challenge in CHORD is managing the risk of how purchased goods and services perform.  Specifically, project management must constantly balance two extremes within a fixed budget: the risk of under performance versus the risk of cost overruns. 

To illustrate the tradeoff between performance and cost (and how risk factors in), it is instructive to consider two broad pathways offered by traditional procurement: component-level sourcing and turnkey manufacturing. There are often many incentives to pursue the latter, such as avoiding the logistical and financial cost of assembly, which is often more efficiently performed by industry. However, this option usually also requires that the vendor assume the risk of the assembly and the performance of the final product. As a result, contract prices for this option can be significantly more than one would assume based on materials and labor alone.

In response, we have adopted a third pathway for some CHORD procurements, which takes advantage of industry's efficiency, while accepting some of the risk we would otherwise be forced to assume under component-level sourcing. Essentially, this pathway only specifies requirements on the component materials, or those that can otherwise be easily met and verified by the vendor. The risks associated with the final assembly, as well as the long-term performance of the delivered goods (which are more difficult for the vendor to guarantee without additional cost) are carried by the project.  In some cases, we may also offer to help the vendor meet certain requirements, by providing tools or assistance at certain stages of the assembly.

These approaches have allowed CHORD to realize significant savings on some purchases, while still managing performance risk to an acceptable level.  However, contracts with these terms are not standard for many institutional procurement offices, and thus can require significant time and effort to enact. Additionally, these approaches are most successful when close relationships are formed with vendors.  These relationships take time to develop, and can represent a risk in themselves by creating over-reliance on a single vendor or reducing competition (an issue also present in Sec.~\ref{sec:procure}).

\subsection{Budget Forecasting}
\label{sec:budget}

The third and final major challenge that we shall discuss here is budget forecasting. As mentioned in Sec.~\ref{sec:pm}, the current pace of technological, political, and economic change can make budget forecasting challenging, even when one has complete knowledge of what is needed.  Supply, demand, and tariffs can drive significant price changes on the timescale of months, not years. Nevertheless, in CHORD we have been relatively lucky, with price hikes and price drops occurring in roughly equal measure, at least in the areas related to the correlator and computing. The CHORD collaboration also benefits from significant legacy knowledge in these areas, meaning that the initial budgets put forth in the original proposal were largely accurate and complete.

The greater challenge, as mentioned in Ref.~\citenum{Hoff2026Pathfinder}, has been the fidelity of the budget forecast related to dish manufacturing. While the dish technology had been demonstrated through a small number of prototypes, the collaboration had no prior experience manufacturing them in large quantities, and thus there were large uncertainties in the budget estimates for the facility, materials, and labor needed to do so\footnote{Note that the CFI does not provide contingency, and there is no obvious path to fill the funding gap between proof of concept and full scale production.}. Unfortunately, unlike the effects of price fluctuations that largely canceled out, the dish budget was underestimated overall, resulting in significant cost overruns. These overruns have been addressed by securing additional funding from other sources.

At the time of writing these proceedings, the remaining uncertainties in the dish manufacturing budget have shrunk considerably and, together with the additional funding secured, we have largely retired the risks associated with this issue. Nevertheless, a clear lesson is that lack of direct experience in the planning stages can lead to underestimation of the associated budgetary risk. Additionally, finding appropriate additional sources of funding and writing proposals to address the budget shortfall has taken considerable time and effort from senior collaboration members and project management.

\section{CONCLUSION}
CHORD represents a case study in navigating the tight constraints of a fixed-budget, fixed-timeline research infrastructure project during a period of unprecedented global economic and technological volatility. By adopting some principles of agile project management, CHORD has proven resilient against this volatility and many of the challenges it presents. In addition, CHORD has implemented modifications to traditional procurement processes, including iterative vendor consultation and tailored approaches to balance risk and cost on contracts. Despite these measures, managing accurate budget forecasts has been a formidable challenge, especially in areas of the project with less legacy knowledge from previous projects.

Barring some early delays resulting from the COVID-19 pandemic, CHORD has remained largely on schedule and on track to deliver the high-level capabilities laid out in the initial proposal.  At the time of writing these proceedings, all major subsystems to enable observations with the first 64-dish sub-array have been delivered to the observatory, and integration is actively underway.  On the timescale of the next few months, we plan to begin commissioning activities with the first 64 dishes, while construction of the remaining dishes and outrigger sites continues in parallel.

\acknowledgments 

The CHORD core array is situated on the ancestral and unceded territory of the Syilx (Okanagan) Nation. The CHORD collaboration recognizes the enduring stewardship of these lands by the Syilx people and are grateful for the opportunity to carry out scientific research in this region.

The CHORD collaboration acknowledges the Dominion Radio Astrophysical Observatory (DRAO), operated by the National Research Council Canada (NRC), as a full institutional partner in the CHORD project and host of the CHORD core array on its radio-protected site near Penticton, British Columbia.

CHORD infrastructure is supported through grants from the Canada Foundation for Innovation (CFI), the provinces of Alberta, Ontario, and Quebec, the Natural Sciences and Engineering Research Council of Canada (NSERC), the Istituto Nazionale di Astrofisica (INAF) in Italy, and participating partner institutions.

Additional support was provided by the Dunlap Institute for Astronomy and Astrophysics at the University of Toronto, which is funded through an endowment established by the David Dunlap family. M.D. is supported by a Canada Research Chair and a NSERC Discovery Grant.

\bibliography{report} 

@inproceedings{white_agile_2008,
	title = {Agile project management: a mandate for the changing business environment},
	url = {https://www.pmi.org/learning/library/agile-project-management-mandate-changing-requirements-7043},
	language = {en},
	urldate = {2026-06-01},
	series = {Paper presented at {PMI} {Global} {Congress} 2008—{North} {America}},
	author = {White, Karen R. J.},
	month = oct,
	year = {2008},
}

@article{vanderlinde_lrp_2019,
  author = {Vanderlinde, K. and others},
  title = {The {Canadian} {Hydrogen} {Observatory} and {Radio-transient} {Detector} ({CHORD})},
  journal = {Long Range Plan 2020 White Paper},
  year = {2019},
  eprint = {arXiv:1911.01777}
}

@misc{noauthor_innovation_2024,
	title = {Innovation {Fund}},
	url = {https://www.innovation.ca/apply-manage-awards/funding-opportunities/innovation-fund},
	language = {en},
	urldate = {2026-06-02},
	journal = {Canada Foundation for Innovation},
	month = mar,
	year = {2024},
    howpublished = "Canada Foundation for Innovation  \url{https://www.innovation.ca/apply-manage-awards/funding-opportunities/innovation-fund}", 
    note = "(Accessed: 2 June 2026)" 
}

@inproceedings{Hoff2026Pathfinder,
  author = {Hoff, Brian and Wulf, Dallas and Islam, Mohammad and Dueck, Tayron and Hellyer, Richard},
  title = {Dish Working Group Progress Toward the {CHORD} 512-Dish Core Array: Completion of the 64-Dish Pathfinder},
  booktitle = {Radio Telescopes, Technologies, and Methods},
  series = {Proceedings of SPIE},
  volume = {14153},
  pages = {14153-62},
  year = {2026}
}

@inproceedings{Dobbs2026,
  author = {Dobbs, Matt and others},
  title = {Overview of the {Canadian} {Hydrogen} {Observatory} and {Radio}-transient {Detector} ({CHORD}) {Project}},
  booktitle = {Radio Telescopes, Technologies, and Methods},
  series = {Proceedings of SPIE},
  volume = {14153},
  pages = {14153-51},
  year = {2026}
}
\bibliographystyle{spiebib} 

\end{document}